\documentclass[a4paper]{jpconf}

\usepackage{graphicx}
\usepackage{amsmath}
\usepackage{amssymb}
\usepackage{epsfig}
\usepackage{color}

\usepackage[colorlinks=true
,urlcolor=blue
,anchorcolor=blue
,citecolor=blue
,filecolor=blue
,linkcolor=blue
,menucolor=blue
,pagecolor=blue
,linktocpage=true
,pdfproducer=medialab
,pdfa=true
]{hyperref}
\usepackage[all]{hypcap}

\begin{document}

\boldmath
\title{NLO QCD corrections to $Wb\,W\bar b$\/ production at hadron
  colliders: new developments and new issues}
\unboldmath

\author{Jan Winter}
\address{Max Planck Institute for Physics, F\"ohringer Ring 6, 80805
  Munich, Germany}
\ead{jwinter@mpp.mpg.de}

\begin{abstract}
This short summary reviews the recent progress in the computation of
higher-order corrections to $W^+b\,W^-\bar b$\/ production. In
addition, new phenomenological studies reveal potential problems that
may affect precision measurements at the LHC.
\end{abstract}

\vspace*{-2mm}
\section{Introduction}
After the first round of LHC running at proton--proton collision
energies of $7$ and $8\mathrm{\:TeV}$, the interest in, and demand for
high precision calculations has increased substantially. The top quark
sector has been in the focus with a large number of publications. In
particular the flagship process, top quark pair production, attracted
a lot of attention. However in the experiments, top quarks can only be
traced via their decay products, which means that for any experimental
selection, the more realistic, and more physical, final state
description will be achieved by computing $W^+b\,W^-\bar b$\/
production directly, cf.~\cite{Krauss:Top2014,Caola:Top2014}. This way
one includes effects that are beyond the ($t\bar t$-like)
approximation of considering only double-resonant top quark
propagators. The contributions from single-resonant ($Wt$-like) and
non-resonant ($VV$-like) diagrams are also taken into account together
with all related quantum interferences resulting from combining the
various contributions. In addition, offshell effects will be well
captured by this approach.

The first higher-order QCD calculations to the hadro-production of
$W^+b\,W^-\bar b$\/ were provided by two different
groups~\cite{Denner:2010jp,Bevilacqua:2010qb,Denner:2012yc} and relied
on the massless $b$\/ quark approximation, i.e.~the five-flavour (5F)
scheme was adopted to accomplish these computations. Furthermore,
finite top quark and $W$\/ width effects were incorporated in a
consistent manner by utilizing the complex mass scheme. Any
preceding work was based on treating the top quarks in the narrow
width approximation (NWA)~\cite{Bernreuther:2004jv,Melnikov:2009dn}.
In this $\Gamma_t\to0$ limit, only double-resonant contributions
survive, and the pair production of onshell top quarks factorizes from
their decays. The neglected contributions can be estimated to be
suppressed by powers of $\Gamma_t/m_t\lesssim1\%$. For sufficiently
inclusive observables and/or kinematical requirements projecting out
the onshell $t\bar t$\/ contributions, small corrections were indeed
seen in a number of comparisons between the predictions of the
{\it full}\/ and {\it factorized}\/
approach~\cite{AlcarazMaestre:2012vp,Denner:2012yc}.
In~\cite{AlcarazMaestre:2012vp}, no more than $1\%$ deviations were
found for inclusive cross sections including experimental cuts.
However, finite width effects can grow (significantly) larger in
differential distributions such as the transverse momentum of the
$b\bar b$\/ pair.

The renewed interest in $Wb\,W\bar b$\/ calculations is twofold and
firmly backed up by the rapid progress in QCD NLO automation. From a 
computational point of view, it is challenging to consistently
incorporate the $b$\/ quark mass effects. From a phenomenological
point of view, one needs to scrutinize whether the NWA can also be
applied to more exclusive phase space regions. In addition, one starts
worrying that at the current experimental precision this whole class
of subleading corrections from offshell effects, non-factorizing
contributions, the $b$\/ quark mass dependence etc.\ has to be
included one way or another. Certainly, over the course of the last
year, new developments in both directions have occurred, and will be
briefly discussed below.

\boldmath
\section{Recent four-flavour scheme calculations and phenomenological
  applications}
\unboldmath

The end of last year has seen two new publications by two independent
groups~\cite{Frederix:2013gra,Cascioli:2013wga} where the finite mass
of $b$\/ quarks has been taken into account in the NLO calculation of
$W^+b\,W^-\bar b$ final states leaving purely leptonic signatures.
Thereby it is important to stress that these predictions have been
obtained for the more complex $pp\to e^+\nu_e\,\mu^-\bar\nu_\mu\,b\bar b$\/
scatterings. The inclusion of finite $m_b$ is a significant step in
gaining better theoretical control of phase space regions with
unresolved $b$~quarks where offshell and single-top contributions are
expected to play a more prominent role.
In the earlier calculations~\cite{Denner:2010jp,Bevilacqua:2010qb,Denner:2012yc}
based on the 5F treatment, the requirement of two hard $b$~jets was
absolutely necessary to produce infrared finite results.
In the four-flavour (4F) scheme however, this does not have to be the
case. Here, a fully differential, NLO accurate description of both
final state $b$\/ jets is achieved, which permits the application of
jet vetoes, and enables one to separate, in a gauge-invariant way, the
narrow width top quark contributions from those of the finite width
remainder. In other words, a unified description of $t\bar t$\/ and
$Wt$\/ production is provided.
Hence, the 4F scheme $Wb\,W\bar b$\/ computation can be used to obtain
reliable estimates in many analyses where top quark pair, $Wt$\/ and
$b$\/ quark associated $WW$\/ contributions constitute an important
background to BSM searches or SM measurements in the electroweak
sector.

\begin{figure}[t!]
  \begin{minipage}[c]{0.39\columnwidth}
    \centering
    \includegraphics[width=\columnwidth]{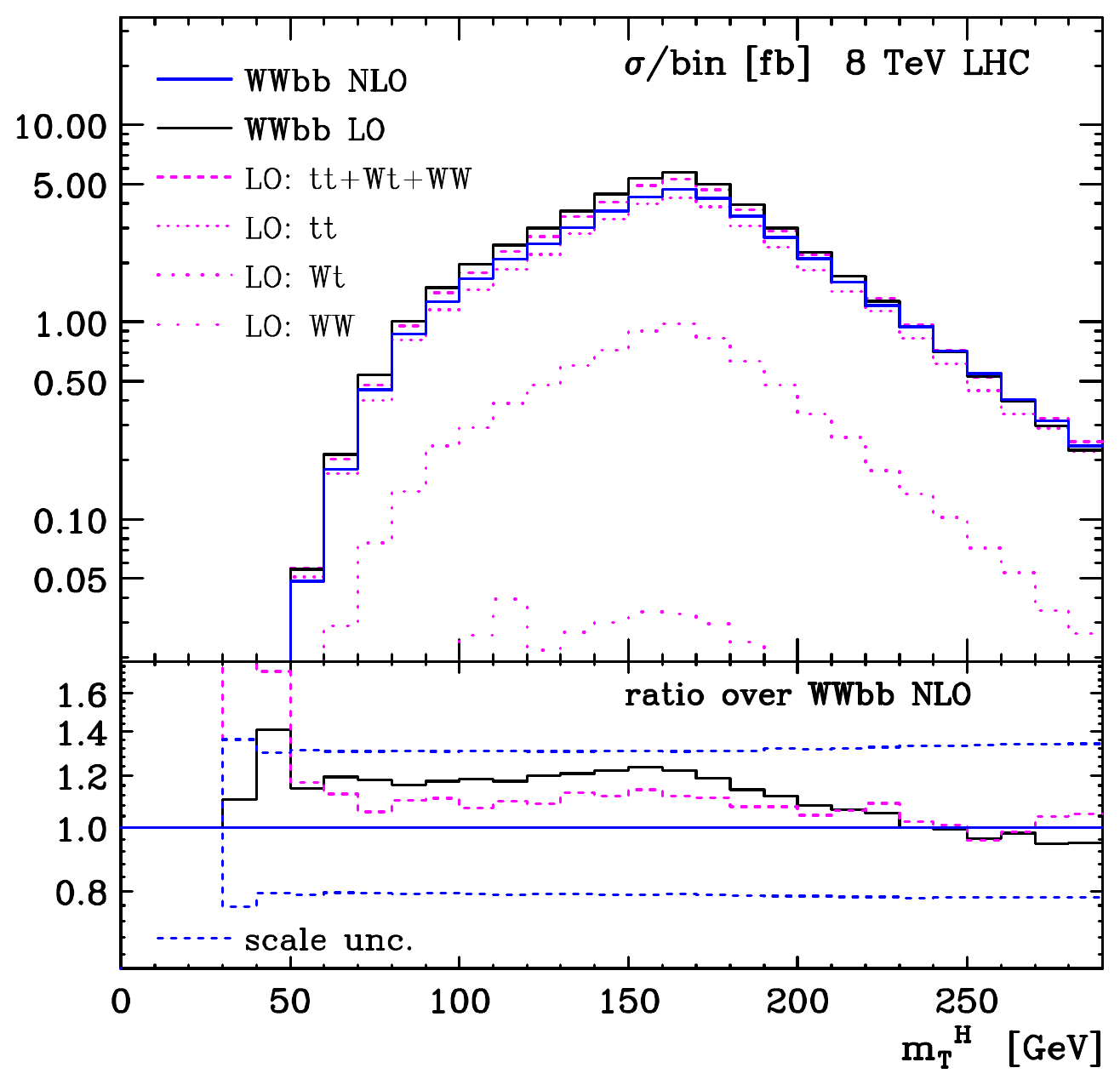}
  \end{minipage}
  \hfill
  \begin{minipage}[c]{0.59\columnwidth}\vskip-8pt
    \caption{\label{fig:frederix}\small%
      Transverse mass distribution of the Higgs boson at NLO/LO
      (blue/black curve) for the $pp\to e^+\nu_e\,\mu^-\bar\nu_\mu\,b\bar b$\/
      process calculated in the 4F scheme. Measurement cuts similar to
      the one-jet bin selection employed by ATLAS have been applied,
      and supplemented by the ``Higgs signal'' topology cuts,
      $m_{ll}<50\mathrm{\:GeV}$ and $|\Delta\phi_{ll}|<1.8$. For the
      4F NLO result, the envelope of a set of scale factor variations
      concerning $(\mu_\mathrm{R},\mu_\mathrm{F})$ is indicated by the
      dashed blue lines. The deviations between the full 4F
      calculations turn out to be larger than those occurring between
      the full 4F prediction at LO and the incoherent sum (dashed red
      line) of NWA predictions at LO for $t\bar t$, $Wt$\/ and
      $b$~quark associated $ll\nu\nu$\/ production. The Higgs boson
      transverse mass is defined via
      $(m^H_T)^2=(E^{ll}_T+E^\mathrm{miss}_T)^2-
      |\vec p^{\;ll}_T+\vec p^\mathrm{\;miss}_T|^2$
      where $(E^{ll}_T)^2=|\vec p^{\;ll}_T|^2+m^2_{ll}$.}
  \end{minipage}
\end{figure}

In the study by R.~Frederix, see Ref.~\cite{Frederix:2013gra}, the
focus is on one example of particular importance to Higgs boson
measurements at the $8\mathrm{\:TeV}$ LHC.
Considering the $WW^{(\ast)}$ channel with fully leptonic decays, the
top quark induced background in the one-jet bin has been evaluated
using the newly available 4F calculation of $Wb\,W\bar b$\/ production.
The parton level computations were performed within the
\textsc{MadGraph5\_aMC@NLO} framework, and the events were required to
pass a one-jet bin selection closely following the actual ATLAS
strategy for this measurement. \Fref{fig:frederix} shows one example
plot of the analysis carried out in~\cite{Frederix:2013gra}, including
the final step of imposing the ``Higgs signal'' topology cuts.


\begin{figure}[t!]
  \begin{minipage}[t]{0.485\columnwidth}
    \centering
    \includegraphics[height=0.78\columnwidth,angle=0]{./figs/1312.0546/%
      1312-0546_njet}
    \caption{\label{fig:jetbins}\small%
      4F scheme $W^+W^-b\bar b$\/ cross sections, $K$-factors and
      finite $\Gamma_t$ contributions in the $0$-jet, $1$-jet and
      inclusive $2$-jet bin, using two different scales.}
  \end{minipage}
  \hfill
  \begin{minipage}[t]{0.485\columnwidth}
    \centering
    \includegraphics[height=0.78\columnwidth,angle=0]{./figs/1312.0546/%
      LHC8-allDS2-XS-distributionCV-cp05KfactorCV-cp03Delta-phi-emmup-new-gp_}
    \caption{\label{fig:dphi0jet}\small%
      $0$-jet bin distribution~of~the~transverse opening angle of
      the charged leptons. The red/\-blue bands indicate scale
      uncertainties at NLO/LO.}
  \end{minipage} 
\end{figure}

The work of F.~Cascioli et al.~provides a detailed discussion of the
impact of NLO and finite (top quark and $W$) width corrections for
many jet bin definitions of interest~\cite{Cascioli:2013wga}.
Particular attention has been paid to the effects of jet $p_T$ vetoes
of different strengths, $n_\mathrm{jet}$ and jet flavour requirements.
The results have been obtained with an in-house NLO parton level
generator using the capabilities of \textsc{OpenLoops+Collier}.
As the ill-defined $t\bar t$/$Wt$\/ separation of the 5F scheme is
avoided, they furthermore introduce a dynamical scale based on
transverse energies $E_{T,i}=\big(m^2_i+p^2_{T,i}\big)^{1/2}$, which
interpolates between $\mu^2_{t\bar t}=E_{T,t}\,E_{T,\bar t}$ and
$\mu^2_{tW^-}=E_{T,t}\,E_{T,\bar b}$ for $t\bar t$\/ and single-$t$\/
topologies, respectively. This better accounts for the multi-scale
problem at hand.
\Fref{fig:jetbins} compares results based on this new scale choice,
$\mu=\mu_{WWb\bar b}$, and the more commonly used scale choice
$\mu=m_t$. The presentation is broken down into different
$n_\mathrm{jet}$ contributions that add up to the fully inclusive
result. This makes it clear that the global ${\cal O}(1.4)$ $K$-factor
emerges mainly from the large inclusive two-jet bin corrections, which
are slightly larger for the computation relying on the dynamical
scale. One observes that the finite top quark width effects are
strongly enhanced for the exclusive one-jet and zero-jet selections.
They may grow as large as $30\%$ if no jets are allowed. This is also
nicely demonstrated in \Fref{fig:dphi0jet} visualizing the
differential distribution of the azimuthal angle separation between
the two leptons in the event. This observable is key to precision
measurements of spin correlations in $t\bar t$\/
production~\cite{ATLAS-CONF-2014-056}, and clearly, a flattening of
the $\phi_{l^+l^-}$ shape as a result of the NLO treatment leads to
effects in determining parameters such as $f_\mathrm{SM}$, which
quantifies how SM-like the spin correlations are~\cite{ATLAS-CONF-2014-056}.

\begin{figure}[b!]
  \begin{minipage}[c]{0.37\columnwidth}
    \centering
    \includegraphics[width=\columnwidth,angle=0]{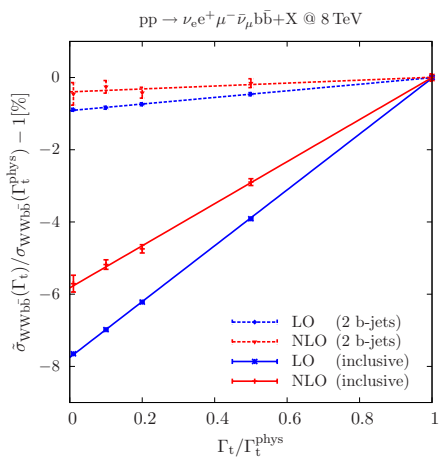}
  \end{minipage}
  \hfill
  \begin{minipage}[c]{0.61\columnwidth}\vskip-7pt
    \centering
    \caption{\label{fig:width-extrapolation}\small%
      Numerical extrapolation to the $\Gamma_t\to0$ limit of the NLO
      (red lines) and LO (blue lines) cross sections for
      $W^+W^-b\bar b$\/ hadro-production.
      The results are presented in terms of relative differences that
      employ the respective 4F scheme cross sections obtained with the
      physical top quark width as their reference.
      The numerical NWA used here is expressed as
      $d\sigma_{t\bar t}=\lim_{\Gamma_t\to0}\limits
      \big(\Gamma_t/\Gamma^\mathrm{phys}_t\big)^2
      d\sigma_{WWb\bar b}(\Gamma_t)$,
      and shown for two jet phase space selections (including leptonic
      cuts): one that reflects the inclusive case (solid lines) and
      one where two $b$~jets have to be resolved (dotted lines).
      While the $t\bar t$\/ signal-like selection shows little
      dependence on finite width corrections, it is crucial to take
      them into account for the inclusive phase space. The finite top
      quark width remainder turns out to be dominated by $Wt$\/
      contributions.}
  \end{minipage} 
\end{figure}

The rather different impact of using a finite $\Gamma_t$ is also
depicted in \Fref{fig:width-extrapolation}. A numerical extrapolation
to the NWA is applied to compare the relative cross section
differences $\sigma_{t\bar t\textrm{-like}}/\sigma_{WWb\bar b}-1$
between a rather inclusive event selection and one requiring two hard
$b$~jets.



\section{Top quark mass determination -- implications
  of a parton level study}
Automated/modern NLO generators provide us with opportunities to not
only explore more intricate phase space regions but also re-examine
our predictions for important observables used in precision studies.
Accordingly, the cutting-edge parton level $W^+b\,W^-\bar b$\/
calculations are a perfect means to help disentangle effects in
$t\bar t$\/ production that go beyond the factorization of top quark
production and subsequent decay. Their phenomenological relevance can
be assessed in comparison to NWA and standard Monte Carlo approaches.
This is of paramount importance to scrutinize approaches that have
been taken to facilitate precision measurements such as the
determination of top quark mass parameters that are (very) closely
related to the top quark pole mass, $m_t$.
Building on a very rich and active experimental program, the first
combination of LHC and Tevatron results on $m_t$ culminated in finding
a total uncertainty on $m_t$ of less than
$1\mathrm{\:GeV}$~\cite{ATLAS:2014wva}. The tension between the
results of individual measurements however is rising.

One recent ATLAS measurement contributing to this combination utilizes
the $m_t$ sensitivity of the shape of the $m_{lb}$ distribution in
dilepton events to determine the top quark mass as
$m_t/\mathrm{GeV}=173.09\pm0.64\,\mbox{(stat)}\pm1.50\,\mbox{(syst)}$%
~\cite{ATLAS-CONF-2013-077}.
The result of this one-dimensional template fit is based on data
collected at the $7\mathrm{\:TeV}$ LHC with an integrated luminosity
of $4.7/\mathrm{fb}$. While the systematic uncertainty is dominated by
that of the jet energy scale, the ATLAS estimate for the theoretical
uncertainty amounts to $0.8\mathrm{\:GeV}$. The success of the
$m_{lb}$ approach therefore depends on a solid understanding of the
theoretical uncertainties associated with the $m_{lb}$ shape.

\begin{figure}[t!]
  \centering
  \includegraphics[width=0.32\columnwidth]{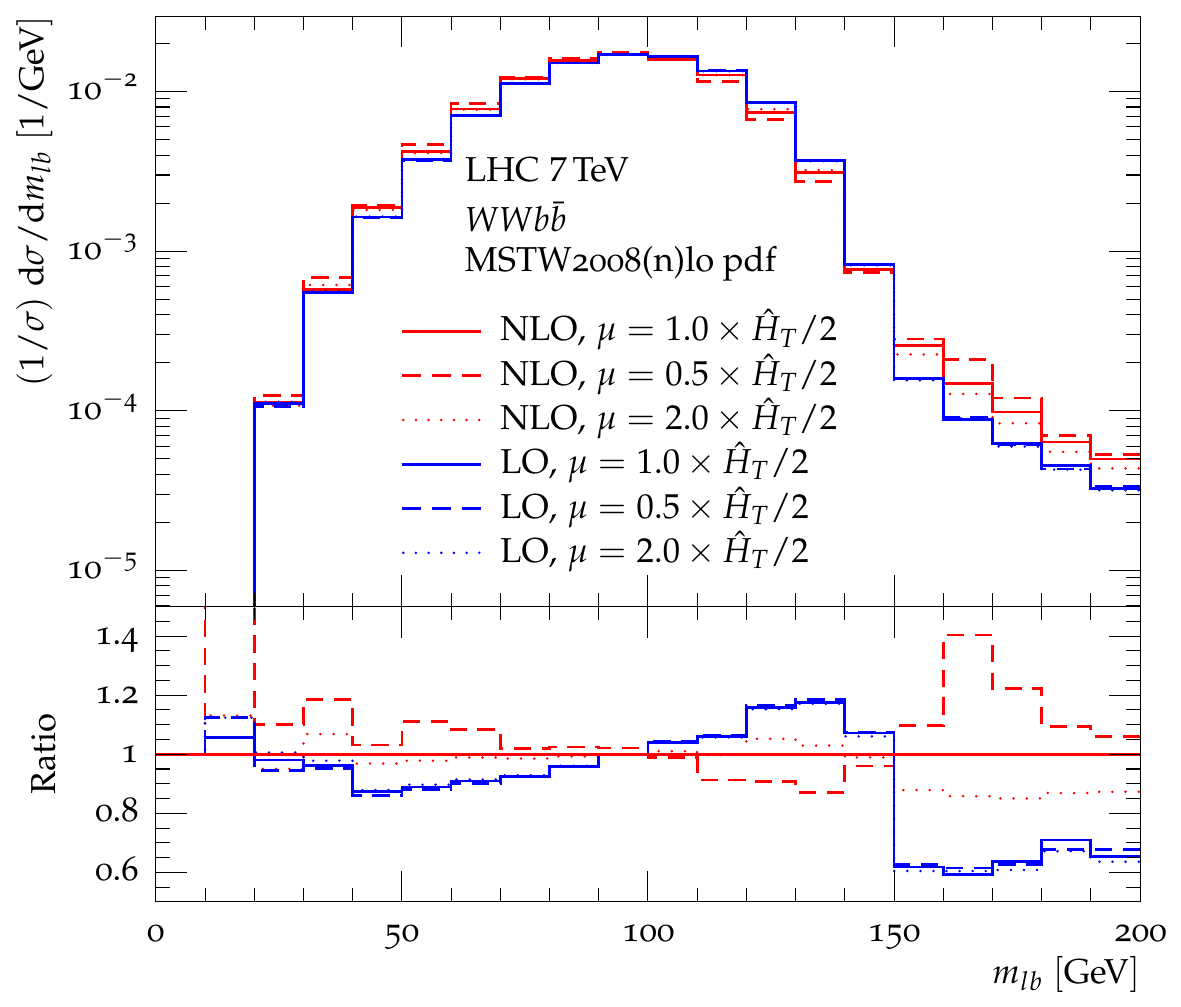}
  \hfill
  \includegraphics[width=0.32\columnwidth]{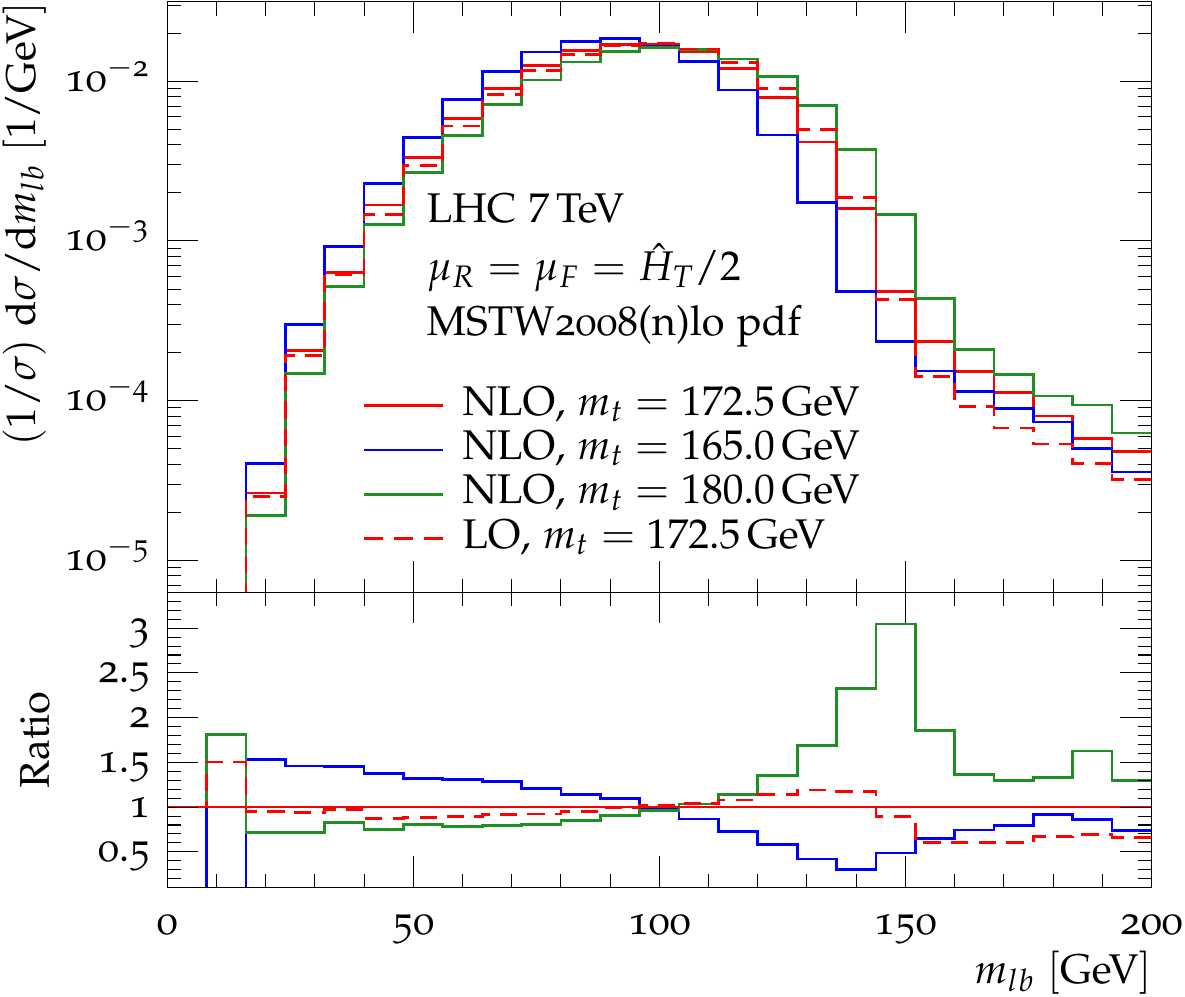}
  \hfill
  \includegraphics[width=0.32\columnwidth]{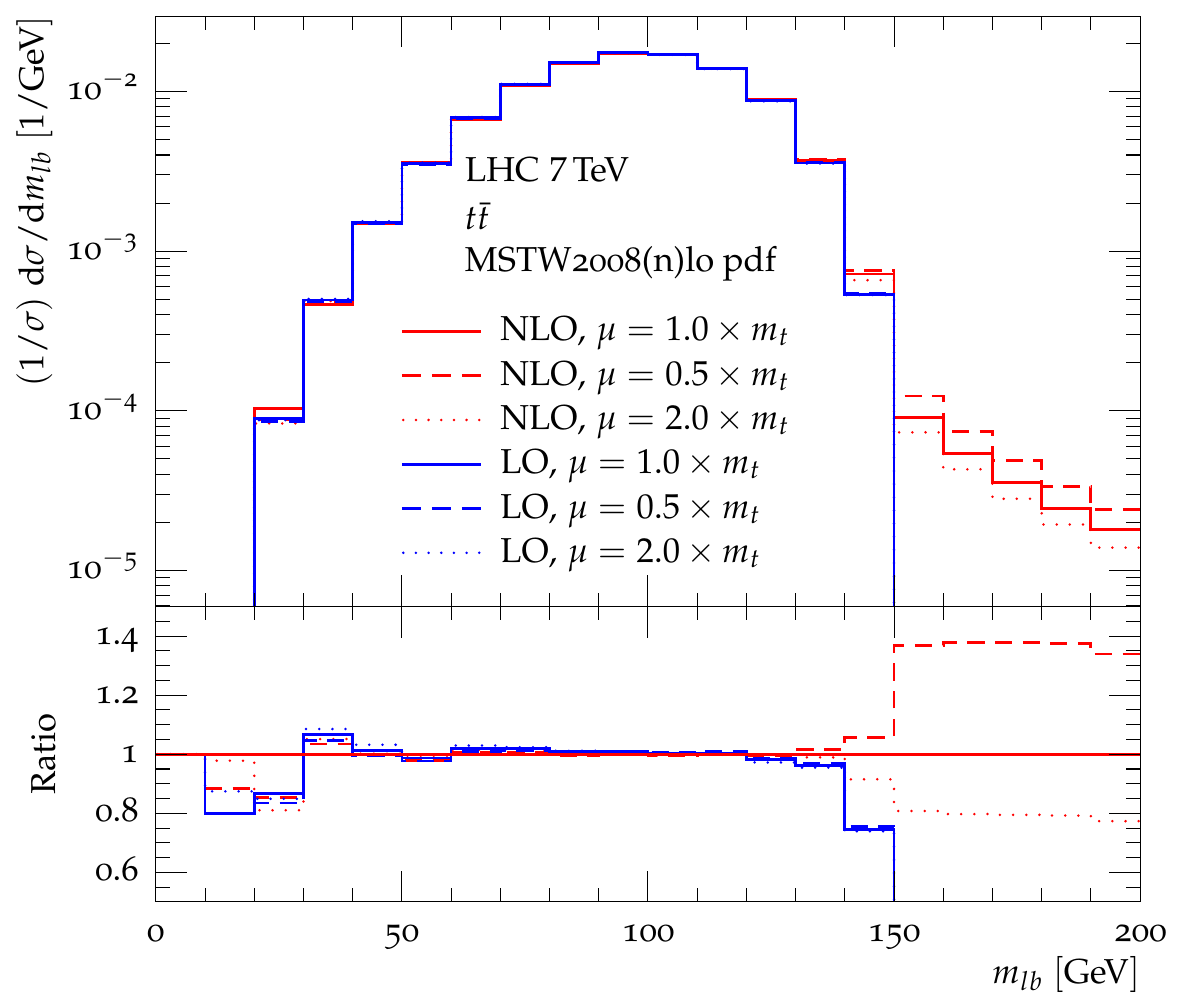}
  \caption{\label{fig:scalevsmtop}\small%
    Various normalized predictions for the parton level $m_{lb}$
    distribution using the calculations of
    Refs.~\cite{Heinrich:2013qaa,Schlenk:2013gza}.
    The outer panels visualize $m_{lb}$ shape changes due to scale
    variations for the full (left) and factorized (right) approach.
    The middle panel shows how different $m_t$ values alter the
    $m_{lb}$ shape as well.}
\end{figure}

Using factor-two scale variations, one can test the robustness of the
$m_{lb}$ shape at the parton level in a straightforward way. In a
factorized, spin correlation preserving approach where the core
$t\bar t$\/ production is described at NLO and supplemented with LO
decays, one obtains very stable results as depicted in the right panel
of \Fref{fig:scalevsmtop}. This should serve as a fairly accurate
model of the current parton level theory standard used in the
experiments which is based on NLO+PS matching as provided for example
by \textsc{PowHeg} or \textsc{MC@NLO}. If one instead uses a full NLO
treatment of the $Wb\,W\bar b$\/ final state as done in
Refs.~\cite{Heinrich:2013qaa,Schlenk:2013gza}, very different, much
more pronounced shape variations are found surprisingly. The results
in the left panel of \Fref{fig:scalevsmtop} originate from a 5F
\textsc{GoSam+Sherpa} computation, and in comparison to the $m_t$
sensitivity of the $m_{lb}$ shape displayed in the middle panel, it is
obvious that the scale factor variations of the full calculation mimic
shape changes as induced by different $m_t$ values. Against usual
expectations, the more accurate theory approach will therefore produce
larger theoretical uncertainties in the determination of the top quark
mass.

\begin{figure}[t!]
  \begin{minipage}[c]{0.68\columnwidth}
    \hskip51mm
    \includegraphics[width=0.5\columnwidth]{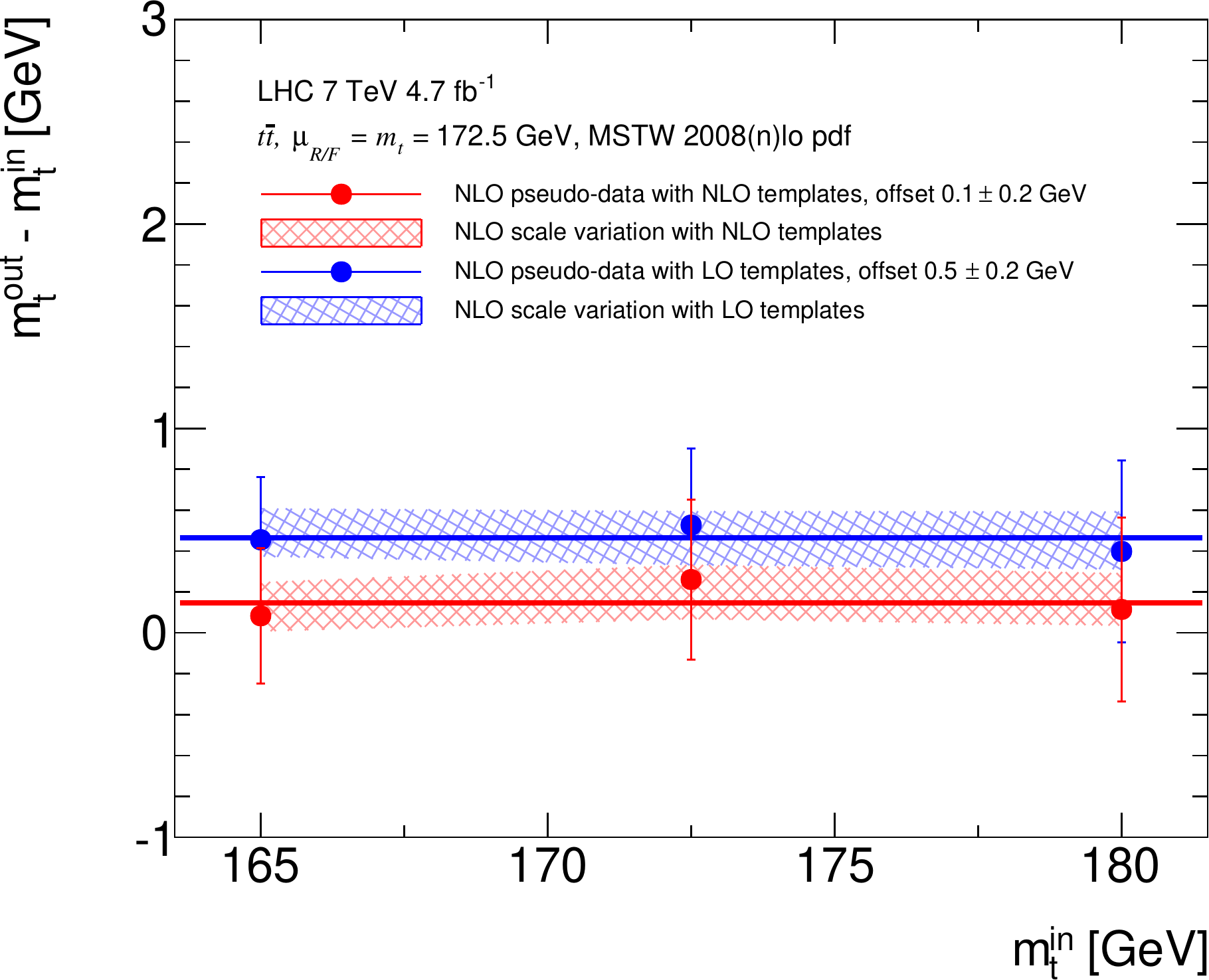}
    \hskip-107mm
    \includegraphics[width=0.5\columnwidth]{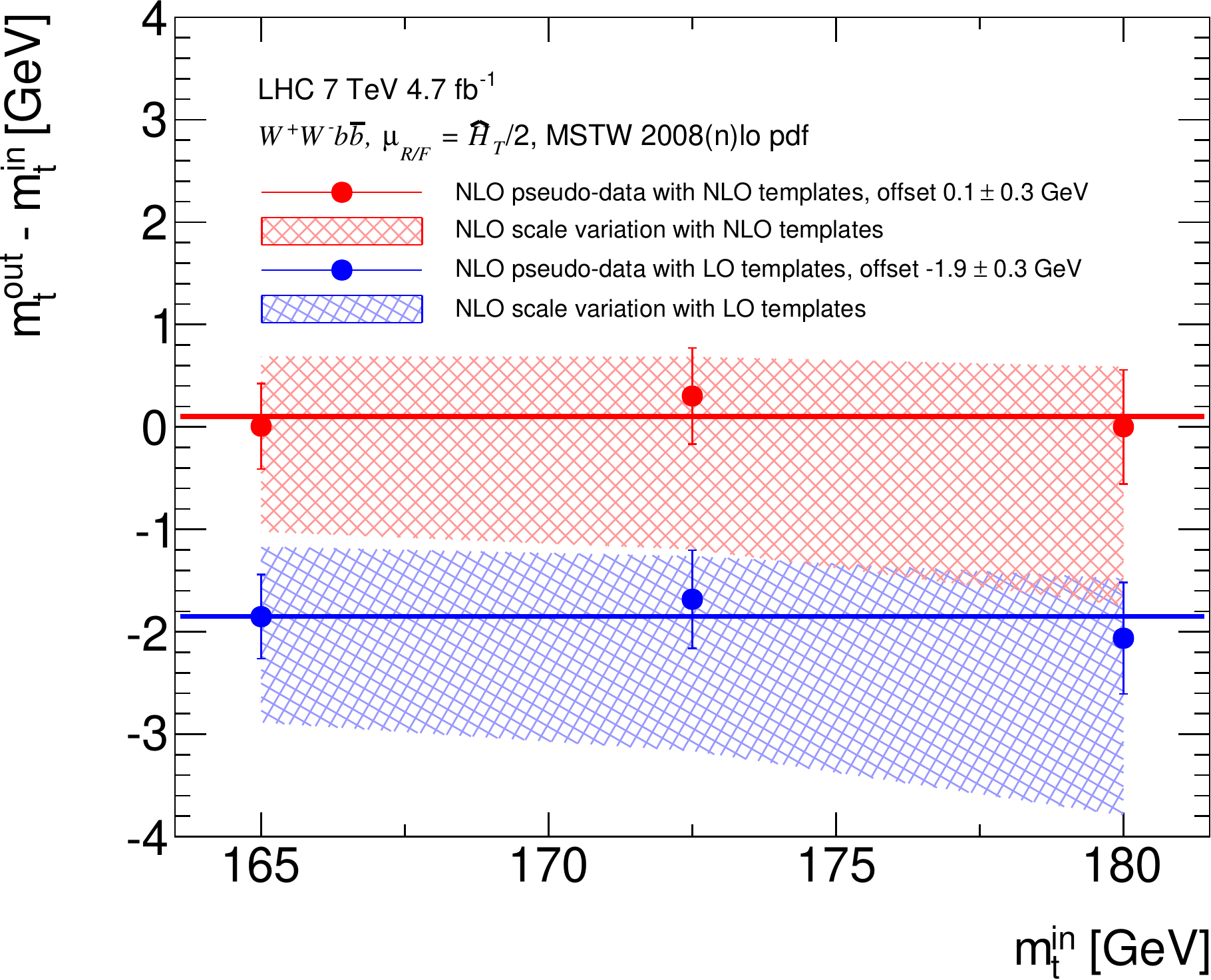}
  \end{minipage}
  \hfill
  \begin{minipage}[c]{0.31\columnwidth}\vskip-10pt
    \caption{\label{fig:redbluebands}\small%
      The $m_t$ offset predictions, for three $m_t$ input values,
      as estimated from various $7\mathrm{\:TeV}$ LHC
      pseudo-experiments of $4.7/\mathrm{fb}$ data luminosity. The
      points show the mean offsets and their statistical uncertainty.
      The bands indicate how the offset varies owing to theoretical
      uncertainties on the $m_{lb}$ shape.}
  \end{minipage}
\end{figure}

This has been demonstrated in a parton level analysis in
Ref.~\cite{Heinrich:2013qaa} where the $m_{lb}$ template fitting
procedure, cf.~\cite{ATLAS-CONF-2013-077}, has been applied to
pseudo-data generated from NLO and LO predictions at given
$m_t^\mathrm{in}$. This has been done separately for both the full and
factorized approach as well as the default scale and scale varied
choices. The different sets of pseudo-data are then tested against the
theory hypotheses defined by the NLO and LO templates one obtains from
the corresponding default scale choice predictions at different top
quark masses. The templates thus parametrize this $m_t$ dependence,
and by varying the sets of pseudo-data and hypotheses one can single
out two effects, the one caused by the NLO corrections and the one
stemming from scale uncertainties.
The outcome of these parton level template fits is summarized in
\Fref{fig:redbluebands} showing the $m_t^\mathrm{out}-m_t^\mathrm{in}$
differences for the full and factorized approach separately in the
left and right panel, respectively. The plots convey a clear message:
the full calculation gives rise to (significantly) larger mass shifts
whether one compares NLO versus LO descriptions (indicated by the
separation of the red and blue horizontal lines) or scale variations
by factors of two (indicated by the red and blue bands).

Of course for the full approach, the fact of larger theory
uncertainties needs to be verified in a more realistic context. This
however requires matching the new calculations to parton showers,
which has not been fully solved yet due to the issue of intermediate
resonances. Nevertheless, some first studies have recently been
presented in the literature~\cite{Garzelli:2014dka,Campbell:2014kua}.

\ack
I would like to thank the organizers of the TOP 2014 workshop for
creating an atmosphere that inspired many interesting discussions.
Furthermore, I want to thank my collaborators on this project,
J.~Schlenk, R.~Nisius, G.~Heinrich and A.~Maier.

\section*{References}
\bibliographystyle{iopart-num}
\bibliography{wwbb}


\end{document}